\documentclass[12pt]{article}
\usepackage[dvips,usenames]{color}
\usepackage[dvips]{graphicx}
\usepackage[letterpaper,dvips,body={8in,11in},vmargin=2cm,hmargin=2cm]{geometry}
\usepackage{xspace}
\usepackage[american]{babel}
\usepackage[latin1]{inputenc}
\usepackage[T1]{fontenc}
\usepackage{ae}
\usepackage{aecompl}
\usepackage[dvipdfm,
            bookmarks     = true,
            breaklinks    = true,
            hypertexnames = false,
            hyperfigures  = true,
            colorlinks    = true,
            urlcolor      = blue]
{hyperref}
\hypersetup{
  pdftitle    = {Schwinger-Dyson Equations},
  pdfauthor   = {Zack Guralnik <zack@het.brown.edu>},
  pdfsubject  = {High Energy Physics - Theory},
  pdfcreator  = {LaTeX2e with hyperref package},
  pdfproducer = {dvipdfm},
  pdfkeywords = {Quantum Field Theory, Lattice, Symmetry Breaking},
  pdfview     = {FitH}
}
\usepackage{curves}
\usepackage{boxedminipage}
\usepackage[nice,tight]{units}
\usepackage{amsmath,amsfonts,amstext,amssymb,amsbsy,amsopn,amsthm,eucal}
\DeclareMathAlphabet{\mathpzc}{T1}{pzc}{m}{it} 
\usepackage{mathrsfs}
\usepackage{wasysym}
\usepackage{dsfont}
\usepackage{lmodern}
\makeatletter
\def\equalsfill{$\m@th\mathord=\mkern-7mu
  \cleaders\hbox{$\!\mathord=\!$}\hfill
  \mkern-7mu\mathord=$}
\makeatother

\def\DHLhksqrt#1#2{\setbox0=\hbox{$#1\sqrt{#2\,}$}\dimen0=\ht0
  \advance\dimen0-0.2\ht0
  \setbox2=\hbox{\vrule height\ht0 depth -\dimen0}%
{\box0\lower0.4pt\box2}}
\newcommand{\ip}[2]{\ensuremath{\langle#1 | #2\rangle}\xspace}
\newcommand{\ipop}[3]{\ensuremath{\langle#1 | #2 | #3\rangle}\xspace}

\newcommand\bg{\begin{eqnarray}}
\newcommand\ed{\end{eqnarray}}
\newcommand\bgn{\begin{eqnarray*}}
\newcommand\edn{\end{eqnarray*}}

\def\del{\partial}

\begin{document}
\begin{titlepage}
  \begin{flushright}
    {BROWN-HET-1487} \\
  \end{flushright}

  \bigskip
  \begin{center}
    {\LARGE \textbf{\textsf{Complexified Path Integrals and the Phases of Quantum field Theory}}} \\
    \vspace{1cm}
    \Large{\textsf{G. Guralnik and Z. Guralnik}} \\
    \bigskip
    \textsf{Physics Department\footnote{\noindent\texttt{WWW:}
        \href{http://chep.het.brown.edu/}{\texttt{chep.het.brown.edu}}\\
        \texttt{Email:} \href{mailto:zack@het.brown.edu}{zack@het.brown.edu},
        \href{mailto:gerry@het.brown.edu}{gerry@het.brown.edu} }} \\
    \textsf{Brown University, Providence --- RI. 02912}
    \date{\today}
  \end{center}

  \bigskip
  \begin{abstract}
    \noindent The path integral by which quantum field theories are
    defined is a particular solution of a set of functional differential
    equations arising from the Schwinger action principle. In fact these
    equations have a multitude of additional solutions which are
    described by integrals over a complexified path. We discuss
    properties of the additional solutions which, although generally
    disregarded, may be physical with known examples including
    spontaneous symmetry breaking and theta vacua. We show that a
    consideration of the full set of solutions yields a description of
    phase transitions in quantum field theories which complements the
    usual description in terms of the accumulation of Lee-Yang zeroes.
    In particular we argue that non-analyticity due to the accumulation
    of Lee-Yang zeros is related to Stokes phenomena and the collapse of
    the solution set in various limits including but not restricted to,
    the thermodynamic limit. A precise demonstration of this relation is
    given in terms of a zero dimensional model. Finally, for zero
    dimensional polynomial actions, we prove that Borel resummation of
    perturbative expansions, with several choices of singularity
    avoiding contours in the complex Borel plane, yield inequivalent
    solutions of the action principle equations.
  \end{abstract}

  \tableofcontents
\end{titlepage}
\section{Introduction}
Many of the ideas presented in this paper have also appeared
previously in \cite{us}. The intent of this article is to assemble
what we view as the important points in a short and coherent
summary, and to add some results concerning a relation between
Lee-Yang zeroes and Stokes phenomena.

Quantum field theories may be defined either by a path integral or
by a set of functional differential equations which follow from the
Schwinger action principle \cite{Schwinger},

\begin{equation}
  \delta \ip{t_1}{t_2} = \ipop{t_1}{\delta S}{t_2} \; .
\end{equation}

For the sake of illustration, consider a zero dimensional ``quantum
field theory'' defined by the action $S(\phi)$. The generating
functional for correlation functions of $\phi$ is

\begin{equation}
  \label{genf}
  Z(J) = \int_{-\infty}^{+\infty} d\phi\, \exp\bigl(-S(\phi) + J\, \phi\bigr) \; ;
\end{equation}
where it is assumed that the integral is convergent. This is a
solution of the action principle equations

\begin{align}
  \label{SD1}
  &\bigl(S'(\partial_J) + J\bigr)\, Z(J) = 0\\ 
  \label{AP1}
  &\biggl(\partial_{g_i} - \frac{\partial S(\partial_J)}{\partial
      g_i}\biggr)\, Z(g,J) =  0\; ;
\end{align} 
where $g_i$ are the parameters of the theory. For the specific action

\begin{equation}
  \label{phifour}
  S(\phi) = \frac{1}{2}\, \mu\, \phi^2 + \frac{g}{4}\, \phi^4 \; ;
\end{equation}
equations \eqref{SD1} and \eqref{AP1} become

\begin{align}
  \label{SD}
  &(g\, \partial_J^3 + \mu\, \partial_J + J)\, Z(J) = 0 \\
  \label{AP}
  &(\partial_g - \frac{1}{4}\, \partial_J^4)\, Z(J) = (\partial_\mu -
    \frac{1}{2}\, \partial_J^2)\, Z(J) = 0 \; .
\end{align}

Although all these equations are included in the
Schwinger action principle, it will be convenient to refer to
equations involving variations of the fields (e.g., \eqref{SD1}) as
Schwinger-Dyson equations, and those arising from variation of
parameters (e.g., \eqref{AP1}) as action principle equations.

In general the equations \eqref{SD1} and \eqref{AP1} have a several
parameter class of solutions. For the action \eqref{phifour}, the
corresponding equations \eqref{SD} and \eqref{AP} have a three
parameter class of solutions, which includes \eqref{genf}. The
Schwinger-Dyson equations are solved by

\begin{equation}
  Z(J) = \sum_I c_I(g,\mu)\, \int_{\Gamma_I} d\phi\, \exp(-S(\phi) + J\, \phi)
    \; ;
\end{equation}
where $\Gamma_I$ are inequivalent integration paths in the complex
$\phi$ plane over which the integral converges and $c_I$ are
arbitrary functions of the coupling constants $g$ and $\mu$. The
number of such contours matches the order of the differential
equation \eqref{SD1}. Figure 1 shows the domains of convergence,
$\cos(4\arg(\phi))>0$, and a basis set of contours for real positive
$g$.

\begin{center}
  \includegraphics{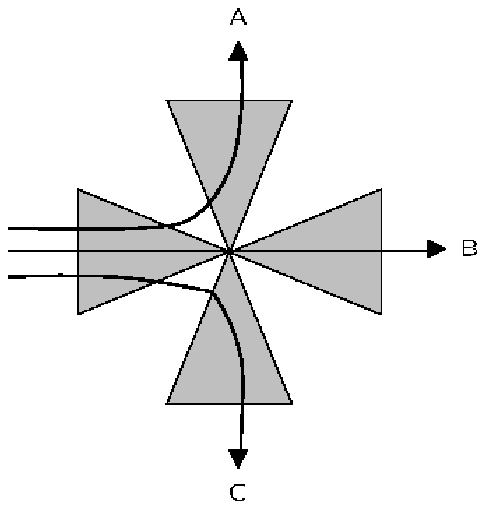}
  \bigskip

  \noindent Figure 1.
\end{center}

Roughly speaking, the action principle requires the coefficients
$c_I$ to be independent of the parameters $\mu$ and $g$. This
statement is imprecise since, if $g$ is is taken to be complex and
the argument of $g$ is varied sufficiently,  the contours of
integration $\Gamma_i$ must be changed to maintain convergence. For
a general polynomial action, this statement holds for the ``top
coupling'' associated with the highest power of $\phi$ in the
action. For a given top coupling $g$, each contour $\Gamma_I$
belongs to an equivalence class of contours for which integration
gives the same result. The equivalence classes (see figure 2)
consist of contours for which $|\phi|\rightarrow\infty$ within the
same domains of convergence; the action is analytic in $\phi$ except
at infinity. There is always a choice of contour $\Gamma_I$ within
an equivalence class which can be held fixed while making
infinitesimal variations of the top coupling. The action principle
requires that the coefficients associated with these contours do not
vary as one makes infinitesimal changes in the couplings.

\begin{center}
  \includegraphics{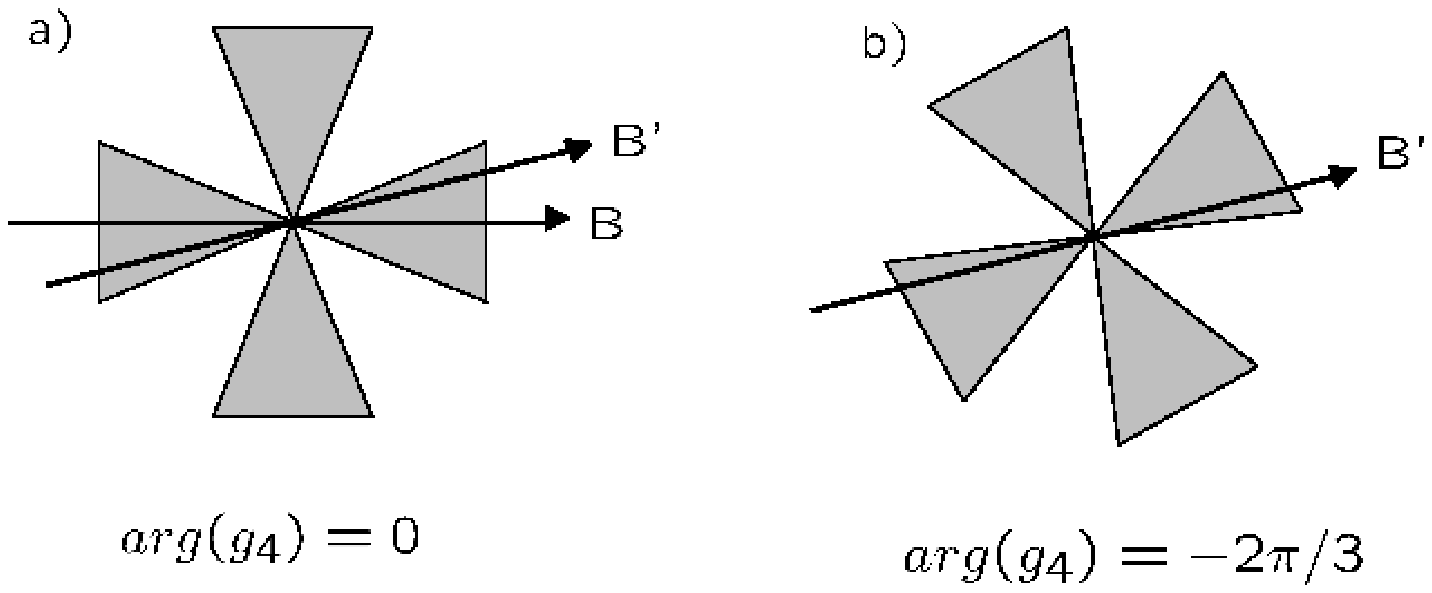}
  \bigskip

  \noindent Figure 2. For $\arg(g_4) = 0$, (a) the contours $B$ and
  $B^\prime$ are equivalent. However for $\arg(g_4) = 2\, \pi/3$, (b) the
  integral over $B^\prime$ remains convergent while the integral over $B$
  diverges.
\end{center}

To dispel doubt that the ``exotic'' solutions with complex
integration contours have physical relevance, consider the effective
action for the theory defined by \eqref{phifour}, with real $\mu \le 0$ and real
$g > 0$. The expectation value for $\phi$ is a solution of the equation

\begin{equation}
  J = \frac{d\Gamma}{d\phi} \; ;
\end{equation}
where $\Gamma$ is the one particle irreducible effective potential. The tree
level effective potential (in this case the action) has 
minima at $\phi = 0,\; \pm \sqrt{\mu/g}$. The minimum  $\langle\phi\rangle =
\sqrt{\mu/g} + \dotsb$ corresponds to a solution of the 
Schwinger-Dyson equations with an integral representation involving
a sum of the contours $A$ and $C$ drawn in figure 1. These integrals
have contributions only from the saddle point at $\phi= \sqrt{-\mu/g}$. Hence
symmetry breaking vacua, whose existence is normally attributed to a
thermodynamic limit, exist even in zero dimensions when the full set of
solutions of the Schwinger-Dyson and action principle equations are considered.
Note that no symmetry breaking term has been added to the action; the symmetry
is broken by the choice of integration path\footnote{The standard way to
non-perturbatively describe symmetry breaking vacua is to introduce
a small symmetry breaking term which is taken to zero only after
taking an infinite volume limit.}.

Note that the solutions associated with the contours $A$ and $C$ in
figure 1. have complex parts which do not appear at any order in a
perturbative expansion about the saddle point. However, due to the
linearity of the Schwinger-Dyson equations, one can sum the contours
to get a solution for which the non-perturbative contribution is
real. The reader might be concerned that the ``exotic'' solutions
which are not integrals over real $\phi$ do not satisfy the axioms
of Euclidean quantum field theory.  Although one can easily arrange
for the Greens functions to be real, one might still worry that even
Greens functions are not manifestly positive. In fact one should
postpone these questions for the higher dimensional case as
perversities of the exotic solutions could vanish in the
thermodynamic/continuum limit. In fact, we expect this to be the
case, as symmetry breaking vacua and theta vacua are examples of
exotic solutions.

In the subsequent section, we argue that the appearance of phase
boundaries in quantum field theories is intimately related to a
collapse of the solution set in the thermodynamic limit.  This
proposal can be made concrete in a zero dimensional analogue, for
which we demonstrate a correspondence between the collapse of the
solution set as a top coupling is set to zero and the accumulation
of Lee-Yang zeros, leading to a non-analyticity in a coupling
constant.  In this context, the limit of a vanishing top coupling is
analogous to the thermodynamic limit. The complementary descriptions
of phase boundaries in terms of the accumulation of Lee-Yang zeroes
and the collapse of the solution set share a common origin in Stokes
phenomena.

In section 3, we prove the equivalence of Borel resummation of the
perturbative expansion about saddle points and exotic solutions of
the Schwinger-Dyson equations, for various singularity avoiding
contours in the Borel plane.  Finally, in section 4, we argue that
the action principle may emerge from the Schwinger-Dyson equations
under suitable conditions, due to the collapse of the solution set
in the thermodynamic limit, rather than being an independent set of
equations.
\section{Collapse of the Solution Set and the Accumulation of Lee-Yang Zeros}
When complex values of the parameters of a quantum field theory are
considered, phase boundaries which appear in the thermodynamic limit
can be understood in terms of the accumulation of zeroes of the
partition function which pinch the real axis,  known as Lee-Yang
zeroes \cite{LeeYang}.  As we shall explicitly demonstrate below, the
accumulation of Lee-Yang zeroes is not confined to the thermodynamic
limit.  In zero dimensional polynomial theories, Lee-Yang zeroes
also accumulate in the limit that the top coupling $g_T$ goes to
zero. In this limit, a non-analyticity in the coupling $g_{T-1}$
appears.  In this context, we will propose an alternative (and
complementary) description of the appearance of phase boundaries in
terms of the collapse of the solution set of the Schwinger-Dyson and
action principle equations.

Consider the zero dimensional action $S= \sum_{l=1}^{T} g_l\, \phi^l$.
There is a branch point at $g_T = 0$, due to the necessity of rotating
the contours of integration in the integral representation, in order
to maintain convergence as the argument of $g_T$ is varied. It is
always possible to find a contour in an equivalence class which can
be held fixed under infinitesimal variations of the top coupling
(see figure 2), so as to satisfy the action principle. However large
variations in the argument of the top coupling require changes in
the contour. In particular, under a rotation by $2\, \pi\, T$ the
solutions transits among $T$ Riemann sheets. The solutions are
analytic in the couplings $g_{l}$ for $l < T$, except at infinity,
since these couplings do not effect the domains of convergence. In
the limit $g_T \rightarrow 0$, the solution develops a branch point
in the new top coupling, $g_{T-1}$. The limit $g_T \rightarrow 0$ is
analogous to a thermodynamic limit. For the case $T = 3$ we will
explicitly show that this limit is accompanied by the accumulation
of Lee-Yang zeroes in the complex $g_{T-1}$ plane.

The appearance of this branch point can be understood in terms of
the collapse of the solution set of the Schwinger-Dyson and action
principle equations.  In the limit $g_T \rightarrow 0$, with fixed
$\arg(g_T)$ solutions of the Schwinger-Dyson and action principle
equations either diverge, vanish or have a finite limit. It is easy
to see which by considering the overlap of the domains of
convergence in the complex $\phi$ plane for the case in which $g_T$
is the top coupling or for which $g_{T-1}$ is the top coupling
($g_T=0$). If a convergent contour for $g_T \ne 0$ is equivalent to
one which lies within a single wedge of convergence for the case
$g_T=0$, then the integral will vanish in the $g_T\rightarrow 0$
limit with fixed $\arg(g_T)$. On the other hand if a convergent
contour for $g_T \ne 0$ is not equivalent to any contour lying
within the wedges of convergence for $g_T = 0$, then the
$g_T\rightarrow 0$ limit is divergent.  A finite $g_T\rightarrow 0$
limit exists only if an equivalent contour lies within two different
wedges of convergence for the $g_T=0$ case. Figure 3 illustrates the
various possibilities for the case $T=4$.

\begin{center}
  \includegraphics{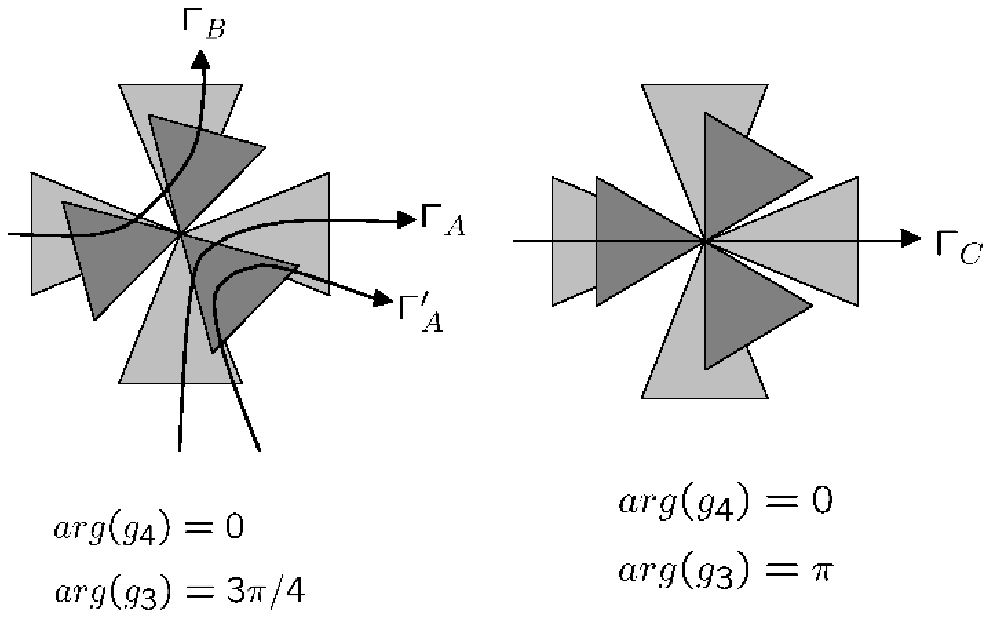}
  \bigskip

  \noindent Figure 3. For $\arg(g_4)=0$ and $\arg(g_3)=3\, \pi/4$, the integral
  over the contour $\Gamma_A$ or the equivalent contour $\Gamma_A^\prime$
  vanishes in the $g_4\rightarrow 0$ limit, while the integral over $\Gamma_B$
  is finite. For $\arg(g_4)=0$ and $\arg(g_3)=\pi$ the integral over $\Gamma_C$
  diverges.
\end{center}

Suppose that the argument of $g_T$ is kept fixed as $g_T \rightarrow
0$.  In this limit the behavior of the partition function, defined
by a particular contour of integration in the complex $\phi$ plane,
will change discontinuously as the argument of $g_{T-1}$ is varied
across certain critical values. For example the $g_T\rightarrow 0$
limit may go from finite to divergent upon crossing a Stokes line.
However it is possible to keep this limit finite by considering
variations in the contour of integration which violate the action
principle by terms which vanish in the $g_T \rightarrow 0$ limit.
For $g_T=0$, these variations are equivalent to analytic
continuation in $g_{T-1}$.

To illustrate how this works, consider the case $T=4$. As one varies
the argument of $g_3$ keeping that of $g_4$ fixed, a generic
solution of the Schwinger-Dyson and action principle equations will
become alternately divergent, finite, or vanishing in the
$g_4\rightarrow 0$ limit.  One can can keep the limit finite by
adding contours such as $\Gamma_A$ in figure 3 when the argument of
$g_3$ enters a wedge in which the contribution from this contour
vanishes as $g_4 \rightarrow 0$.

This process is illustrated in figure 4. The domains of convergence
for $g_4 \ne 0$ are indicated in light gray, while the domains of
convergence for $g_4=0, g_3\ne 0$ are indicated in dark gray.
Starting with a solution of the Schwinger-Dyson and action principle
equations corresponding the contour $A$, $\arg(g_3)$ is varied from
$\pi$ to $\pi/2$ keeping $arg(g_4) =0$. Initially the solution is
finite as $g_4\rightarrow 0$, since the contour $A$ lies
asymptotically within two domains of convergence for $g_4 = 0$. At
$arg(g_3) = 3\, \pi/4$ the contour is modified to $C \equiv A+B$. Note
that the integration over $B$ vanishes in the $g_4 \rightarrow 0$
limit for a neighborhood of $\arg(g_3) = 3\,\pi/4$, since $B$ is
equivalent to a contour lying within a single domain of convergence
when $g_4=0$. One can continue varying $\arg(g_3)$ to $\pi/2$ without
a change in the asymptotic behavior as $g_4\rightarrow 0$; the
partition function remains finite in this limit. On the other hand,
had the contour been fixed as $A$,  the integral would be divergent
as $g_4\rightarrow 0$ for $\arg(g_3)$ on the other side of the Stokes
line $\arg(g_3) = 5\,\pi/8$. For $\arg(g_3)=\pi/2$ and $g_4=0$, the
contour $A$ does not lie within the domains of convergence
asymptotically, unlike the the contour $C$.

\begin{center}
  \includegraphics{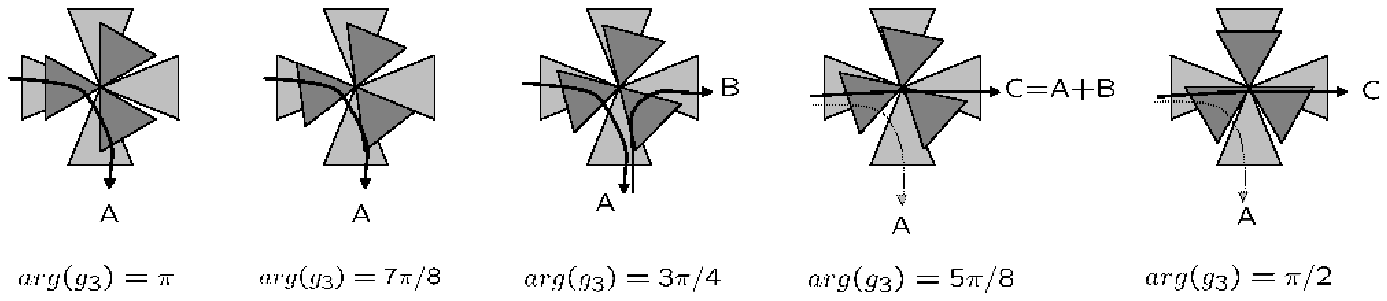}
  \bigskip

  \noindent Figure 4. As the argument of $g_3$ is varied, the contour is changed
  to keep the integral finite in the $g_4\rightarrow 0$ limit.
\end{center}

Considering larger variations of $\arg(g_3)$ one can repeat this process to give
a solution of the Schwinger-Dyson equations which is finite and satisfies the
action principle in the $g_4 \rightarrow 0$ limit. For $g_4 = 0$ this process
corresponds to analytic continuation in $g_{3}$.  The changes in contour
necessary to maintain a finite $g_4\rightarrow 0$ limit (and avoid Stokes
phenomena) give rise to the third order branch point at $g_3 = 0$.

We thus arrive at the picture that non-analyticity in a coupling
constant arises due to the collapse of the solution set of the
Schwinger-Dyson and action principle equations.  In the example we
have given,  non-analyticity in $g_{T-1}$ arises as the top coupling
$g_T \rightarrow 0$.  In this limit, some solutions are finite,
while others diverge or vanish.  Because the equations are linear
and certain solutions vanish in this limit, various classes of
solutions with inequivalent integral representations also coalesce
as $g_T\rightarrow 0$. This permits larger variations of the contour
which satisfy the action principle. At the same time larger
variations of the contour may become necessary to obtain a finite
partition function as coupling constants are varied.  While we have
explicitly demonstrated this mechanism for the $g_T\rightarrow 0$
limit of a zero dimensional polynomial theory, we propose that the
same phenomena hold true for the thermodynamic limit,
$N\rightarrow\infty$ where $N$ is the number of degrees of freedom,
in a multi-dimensional field theory. In other words, non-analyticity
in a parameter of a quantum field theory should be related to the
collapse of the solution set of the Schwinger-Dyson and action
principle equations in the $N\rightarrow\infty$ limit.

The analogy we have proposed between a non-analyticity arising in
the $g_T\rightarrow 0$ limit of a zero dimensional theory and
non-analyticities arising in the thermodynamic limit is strengthened
by noting that the $g_T\rightarrow 0$ limit is accompanied by the
accumulation of Lee-Yang zeroes along Stokes lines in the complex
$g_{T-1}$ plane. One can explicitly see how this occurs for  $T=3$.
Consider the partition function

\begin{equation}
  \label{aint}
  Z = \int_C d\phi\, e^{-(\frac{g}{3}\, \phi^3 + \frac{\mu}{2}\, \phi^2)} \, ;
\end{equation}
for positive real $g$ and a contour $C$ which goes to infinity with
$\arg(\phi) = \pm 2\, \pi/3$, as in figure 5. The integral \eqref{aint}
can be evaluated in terms of an Airy function;

\begin{equation}
  \label{airy}
  Z = 2\, \pi\, i\, e^{-\frac{1}{12}\, \frac{\mu^3}{g^2}}\, g^{-1/3}\,
    Ai(\frac{\mu^2}{4\, g^{4/3}}) \; .
\end{equation}

\begin{center}
  \includegraphics{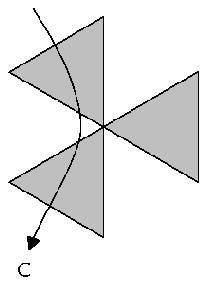}
  \bigskip

  \noindent Figure 5.
\end{center}

The zeroes of $Ai(z)$ lie along the negative real $z$ axis. Since the argument
of the Airy function in \eqref{airy} is $z= \mu^2/4\, g^{4/3}$,  zeroes of the
partition function accumulate on the imaginary axis in the complex $\mu$ plane
as $g\rightarrow 0$, pinching the real axis at $\mu=0$. In fact, $\arg(\mu) =
\pm \pi/2$ are Stokes lines. The $g\rightarrow 0$ behavior of $Z$ in the
neighborhood of $arg(\mu) = -\pi/2$ can be seen by inspecting figure 6, in which
the domains of convergence for $g\ne 0$ and $g = 0$ are indicated by the lighter
and darker shaded regions respectively. For $\arg(\mu) = -\pi/2 + \epsilon$, $Z$
diverges in the $g\rightarrow 0$ limit, with the leading term in a saddle point
expansion given by 

\begin{equation}
  \label{saddle1}
  Z \sim \sqrt{\frac{2\pi}{\mu}}\exp(-\frac{1}{6}\, \frac{\mu^3}{g^2}) \; .
\end{equation}

On the other side of the Stokes line, $\arg(\mu) = -\pi/2 - \epsilon$,
$Z$ converges in the $g\rightarrow$ limit, with the leading term in a
saddle point expansion given by

\begin{equation}
  Z \sim \sqrt{\frac{2\pi}{\mu}}\label{saddle2}
\end{equation}

\begin{center}
  \includegraphics{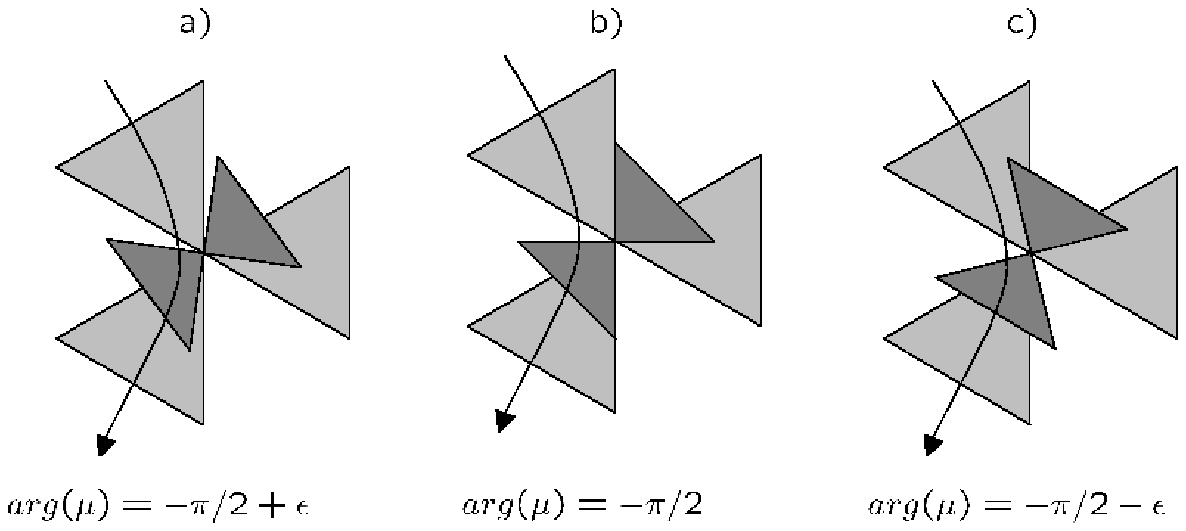}
  \bigskip

  \noindent Figure 6. On either side of the Stokes line at
  $\arg(\mu)= -\pi/2$, the integral is either finite (c) or divergent
  (a) in the $g\rightarrow 0$ limit. Zeroes of the partition function
  accumulate as $g\rightarrow 0$ for values of $\mu$ on the Stokes
  line (b). 
\end{center}

On the Stokes line, $\arg(\mu) =-\pi/2$, the two saddle
point expansions \eqref{saddle1} and \eqref{saddle2} become
comparable in magnitude as $g\rightarrow 0$, since the real part of
the exponential in \eqref{saddle1} vanishes. The integral over the
contour in figure 5 is equivalent to the sum of the integrals over
two constant phase (steepest descent) contours, $C_1$ and $C_2$ in
figure 7, which pass through classical solutions for which the real
part of the action is degenerate. The accumulation of zeroes on the
Stokes line is related to the fact that the relative phase of the
two integrals oscillates wildly with variations of $\mu$ in the
$g\rightarrow 0$ limit, due to the factor of $\exp(\mu^3/g^2) =
\exp(i\, |\mu|^3/g^2)$.

\begin{center}
  \includegraphics{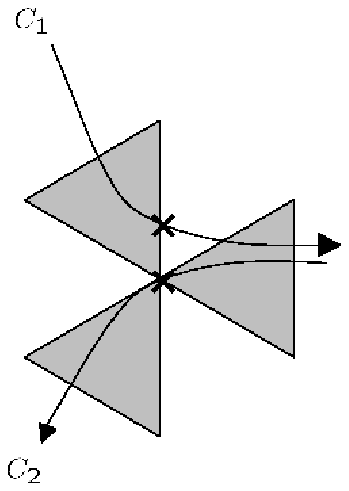}
  \bigskip

  \noindent Figure 7.  Steepest descent contours passing
  through the classical solutions at $\phi = 0,\; -\mu/g$, on the Stokes
  line $\arg(\mu) = -\pi/2,\; \arg(g) = 0$.
\end{center}

Thus, the accumulation of Lee-Yang zeroes and the collapse of the
solution set of the Schwinger-Dyson and action principle equations
are complementary descriptions of the appearance of non-analyticity
in a parameter of a zero dimensional theory as the top coupling
$g_T\rightarrow 0$. The two descriptions have a common origin in
Stokes phenomena. We conjecture that these are also complementary
descriptions of non-analyticity in parameters of quantum field
theories which arise in the thermodynamic limit
$N\rightarrow\infty$. Upon completion of this work, we discovered
\cite{Pisani}, in which Lee-Yang zeroes in a 1D model with a wetting
transition were shown to lie along Stokes lines associated with the
asymptotic expansion in $N$. The correspondence between Stokes lines
and Lee-Yang zeroes has also been suggested in \cite{Itzykson}, in
the context of Ising and gauge models.
\section{Borel Resummation and Complexified Path Integrals}
It is well known that the loop expansion in quantum field theory is
an asymptotic rather than a convergent series.  One approach to
obtain non-perturbative information from the loop expansion is the
Borel resummation, in which a convergent series (the Borel
transform) in a variable $t$ is obtained from the asymptotic series
in $\hbar$.  The Borel transform is then inverted to give a function
having the correct asymptotic expansion, but which contains
non-perturbative information. Since there are actually an infinite
number of such functions, differing by essential singularities at
$\hbar \rightarrow 0$, it is a non-trivial statement that the Borel
transformation corresponds to the path integral. Moreover, there are
frequently singularities of the Borel transform on the positive real
$t$ axis, due to instantons and renormalons, which prevents an
unambiguous Borel resummation.

Under generic conditions, the number of classical solutions is equal
to the number of independent solutions of the Schwinger-Dyson
equations. For an arbitrary polynomial action in a zero dimensions, 

\begin{equation}
  S(\phi) = \sum_{n=1}^T \frac{g_n}{n}\, \phi^n \; ;
\end{equation}
we will prove that various Borel resummations of perturbative expansions about
classical solutions satisfy both the Schwinger-Dyson equations and the action
principle equations, and therefore correspond to various complexified path
integrals.

Consider the partition function

\begin{equation}
  \label{loopZ}
  Z = \int_\Gamma d\phi\, e^{\frac{1}{\hbar}\, S(\phi)} \; ;
\end{equation}
where the path $\Gamma$ is equivalent to a steepest descent path passing through
a dominant saddle point (classical solution) $\phi =\bar\phi_{\alpha}$. The loop
expansion yields a contribution to the generating function of the form;

\begin{equation}
  \label{works}
  Z_{\alpha} \approx \sqrt{ \frac{\pi \hbar}{S^{\prime\prime} (
      \bar\phi_{\alpha} )} }\, e^{- \frac{1}{\hbar}\, S(\bar\phi_{\alpha})}
  \sum_{n=0}^{\infty} c_n \hbar^n \; .
\end{equation}

This series is asymptotic,  but its Borel transform defined by

\begin{equation}
  \label{relbor}
  B_{\alpha}(t) = \sqrt{ \frac{\pi}{S^{\prime\prime} (\bar\phi_{\alpha} ) } }
    \sum_{n=0}^{\infty} \frac{c_n}{\Gamma (n+\frac{1}{2} ) } t^n\; ; 
\end{equation}
converges to

\begin{equation}
  \label{subrl}
  B_{\alpha}(t) = \frac{\sqrt{t}}{2\pi i} \oint_C d\phi\frac{1}{t - ( S(\phi)-
    S(\bar\phi_{\alpha}) ) } \; ;
\end{equation}
where in the vicinity of $t=0$ the contour $C$ encloses, in the opposite sense
(see figure 8), the two poles $\phi_{\alpha,j} (t)$, for $j=1,2$, which coalesce
to $\bar\phi_{\alpha}$ as $t\rightarrow 0$. All the other poles are taken to lie
outside the contour. The Borel transform has a singularity when one of the
exterior poles coalesces with one of the interior poles, which occurs when
$t=S(\bar\phi_{\alpha'}) - S(\bar\phi_{\alpha})$ where $\bar\phi_{\alpha'}$ is a
neighboring classical solution. Doing the $\phi$ integral gives 

\begin{equation}
  \label{als}
  B_{\alpha}(t)= \sqrt{t}\sum_{j=1,2} (-1)^j \frac{1}{S^{\prime}(\phi_{\alpha,j}(t))} \; .
\end{equation}

\begin{center}
  \includegraphics{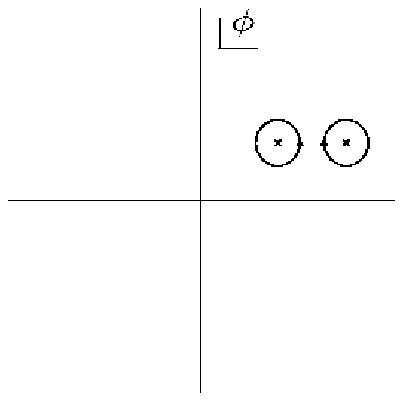}
  \bigskip

  \noindent{Figure 8.}
\end{center}

Thus far everything we have said is well known \cite{Bor}. We now
exhibit an exact relation between the Borel resummation and the
exotic solutions of the Schwinger-Dyson and action principle
equations. We invert the Borel transform by writing 

\begin{equation}
  \label{sqig}
  Z_{\alpha} = e^{-\frac{1}{h}S(\bar\phi_{\alpha}) } \int_\Sigma \, dt\, e^{
    -\frac{t}{\hbar } } \frac{ B_{\alpha}(t)}{\sqrt{t} }
\end{equation}
with an as yet unspecified integration contour $\Sigma$ in the complex $t$
plane.  Equivalence of \eqref{sqig} with a solution of the Schwinger-Dyson
equations requires

\begin{equation}
  \label{equival}
  e^{-\frac{1}{h}S(\bar\phi_{\alpha}) } \int_\Sigma \, dt\, e^{ -frac{t}{\hbar }
  } \oint_C d\phi\frac{1}{t - ( S(\phi)- S(\bar\phi_{\alpha}) ) } = c_I
  \int_{\Gamma_I} d\phi \oint_{\infty}\, dt\, e^{-t/\hbar} \frac{1}{t-S(\phi)}
\end{equation} 
where $\Gamma_I$  indicate open paths in the complex $\phi$ plane which
asymptotically lie within the domains of convergence determined by the the top
coupling $g_T$. Since $C$ and $c_I\Gamma_I$ are inequivalent, \eqref{equival}
involves a non-trivial exchange in the order of integration under which contours
are not preserved. Instead of directly proving \eqref{equival}, we will show
that \eqref{sqig} solves the Schwinger-Dyson equations and satisfies the action
principle.

We set $\hbar  = 1$ in what follows. If $Z_{\alpha}$ satisfies both
the Schwinger-Dyson equations and the action principle then it is
annihilated by the operators;

\begin{equation}
  \label{sdandsac} 
  \hat L = \sum_{n=2}^T (n-1)g_n \frac{\del}{\del_{g_{n-1}}} - g_1 \; ;
\end{equation} 
and 

\begin{equation} 
  \label{sdic}
  \hat H_n = \frac{\del}{\del g_n} - \frac{1}{n}\, \frac{\del^n}{\del g_1^n} \; .
\end{equation} 

It is convenient to define the quantity 

\begin{equation} 
  \label{fdef}
  F_{\alpha} \equiv \int_\Sigma dt e^{-t} \frac{ B_{\alpha}(t)}{\sqrt{t} } =
  \int_\Sigma dt e^{-t}  \sum_{j=1,2} (-1)^j \frac{1}{S^{\prime}(\phi_{\alpha,j})} \; .
\end{equation} 

To show that $\hat L Z_{\alpha} = 0$,  for $Z_{\alpha}$ defined in \eqref{sqig},
it suffices to show that $ \hat{\cal L} F_{\alpha} =0$ where

\begin{equation}
  \label{covar} 
  \hat{\cal L} \equiv \sum_{n=2}(n-1)g_n D_{g_{n-1}} - g_1 \; ; \qquad
  D_{g_{n}} \equiv \frac{\del}{\del g_{n}} - \frac{\del}{\del g_{n}}S(\bar\phi_\alpha) \; . 
\end{equation}

Before proceeding, we list several simple but useful identities. Due
to the equation of motion, $S^{\prime} (\bar\phi_{\alpha} ) = 0$,
one has 

\begin{equation}
  \label{erg} 
  \frac{\del}{\del g_n}S(\bar\phi_{\alpha}) = \frac{1}{n}{\bar\phi_{\alpha}}^n 
\end{equation}

Identities obtained by differentiating the relation $t - S(\phi_{\alpha,j}(t)) +
S(\bar\phi_\alpha)=0$, which defines $\phi_{\alpha,j}(t)$, are 

\begin{equation}
  \label{jig}
  \frac{\del}{\del t}\phi_{\alpha,j}= \frac{1}{S^{\prime}(\phi_{\alpha,j} ) } 
\end{equation}

\begin{equation} 
  \label{wuc} 
  \frac{\del}{\del g_n}\phi_{\alpha,j} = \frac{ \frac{1}{n}(\phi^n_{\alpha,j} -
    \bar\phi^n_{\alpha} )}{S^{\prime} ( \phi_{\alpha,j} )  } \; ;
\end{equation}
where the equations of motion have been used again in deriving the last equation.

Using these identities and the equations of motion, one can show
that 

\begin{equation}
  \label{qant}
  \sum_{n=2}^T [(n-1)g_n D_{g_{n-1}} - g_1]\frac{1}{S^\prime(\phi_{\alpha,j})}
    =0 \; ;
\end{equation}
implying $\hat{\cal L} F_\alpha =0$, or $\hat{\cal L} Z_\alpha =0$, for any
integration path in $t$.

We are not done yet, since the equation ${\hat{\cal L}}Z_\alpha=0$
is a combination of the Schwinger-Dyson and the action principle
equations. The integration path $\Sigma$ will be constrained further
by requiring the action principle to be separately satisfied. To
this end, consider the quantity

\begin{align}
  \nonumber
  {\cal A}&\equiv [klD_{g_l}D_{g_k} + (k+l)D_{g_ {k+l} }]F \\
  \label{yip}
  &= \int_\Sigma dt e^{-t} \frac{\del}{\del t} \sum_j(-1)^j[klD_{g_l}D_{g_k} +
  (k+l)D_{g_ {k+l} }]\phi_{\alpha,j}(t) 
\end{align}
which will vanish if the action principle is also satisfied. Using
the same identities discussed above, ${\cal A}$ may rewritten as

\begin{equation}
  \label{hur}
  {\cal A} = \int_\Sigma dt \frac{\del}{\del t} \left[ e^{-t} \sum_j (-1)^j
    kl \frac{\del}{\del g_k}\frac{\del}{\del g_l}\phi_{\alpha,j}(t) \right] =
  \left. \sum_j (-1)^j e^{-t} kl\frac{\del}{\del g_k} \frac{\del}{\del
      g_l}\phi_{\alpha,j}(t) \right|_{\del\Gamma} \; .
\end{equation}

Clearly ${\cal A}$ vanishes if the contour $\Sigma$ begins at $t=0$ and ends at
$Re(t) = +\infty$, avoiding singularities.  The contribution from the
boundary at $t=0$ vanishes because $\phi_{\alpha,1}(t)$ and
$\phi_{\alpha,2}(t)$ coalesce at $t=0$, so that the factor of
$\sum_j (-1)^j$ in \eqref{hur} leads to a cancellation.  Closed
contours encircling branch cuts in the complex $t$ plane also give
${\cal A}=0$. So long as a contour $\Sigma$ for which ${\cal A}=0$
is used, the Borel resummation gives a solution of the
Schwinger-Dyson and action principle equations.

It would be very interesting to see to what extent this analysis
extends to theories with non-zero dimension.  The analysis is surely
more difficult since, among other possible complications, there are
singularities in the Borel plane due to renormalons as well as
finite action classical solutions (instantons).

The ``exotic'' solutions of the Schwinger-Dyson equations given by
$Z = c_I Z_I$,  where the $Z_I$ are generated from classical
solutions by Borel resummation, may in some sense be thought of as a
generalized form of theta vacua, in which the $c_I$ play the role of
theta parameters. In the usual approach a particular theta vacuum is
selected by adjusting a surface term in the action and integrating
over real fields.  The surface term term effects the Schwinger-Dyson
equations only at the space-time boundary,  so in the infinite
volume limit its role is only to set a boundary condition.  It does
this by putting a different weight on the contributions to $Z$
coming from perturbative expansions about different classical
solutions.  We conjecture that when appropriately resummed, this is
equivalent to a weighted sums over complexified path integrals.
\section{Emergence of Action Principle in the Thermodynamic Limit}
So far we have treated the Schwinger-Dyson and action principle
equations as independent.  This is certainly true for a finite
number of degrees of freedom. If the action principle is not
imposed, the manner in which a solution of the Schwinger-Dyson
equations changes as one varies a coupling is un-determined, as the
there is a continuous set of solutions to the Schwinger-Dyson
equations for any value of the coupling.  However, with certain
assumptions, the action principle arises due to the collapse of the
solution set of the Schwinger-Dyson equations in the thermodynamic
limit.

Suppose that the solution set collapses in the thermodynamic limit,
i.e. some solutions coalesce while others diverge, such that are
solutions of the form $Z \sim \exp(-N {\cal F})$ as
$N\rightarrow\infty$ where there are only discrete possibilities for
the free energy ${\cal F}$. Discreteness of the solutions
automatically fixes the dependence on the couplings. Let us assume
that this dependence is described by a differential equation of the form 

\begin{equation}
  \label{opdef}
  \hat O _{\alpha} Z \equiv ( \frac{\del}{\del g_{\alpha} } - \hat K_{\alpha}
    )\, Z[g_\alpha,J_i]=0 \; ; 
\end{equation}
where $\hat K$ is a linear operator\footnote{The operator $\hat K_\alpha$ is
  necessarily linear for this equation to make sense in the thermodynamic
  limit.}. We will show below that these equations are equivalent to the action
principle equations. The argument follows from the commutation relations of
operators associated with the action principle and the Schwinger-Dyson
equations.

Writing the action in the form $S\{\phi_i\} =  g_\alpha
f_\alpha\{\phi_i\}$, the operators associated with the
Schwinger-Dyson equations are

\begin{equation}
  \hat L_i  \equiv g_\alpha \left.\frac{\partial f_\alpha}{\partial
      \phi_i}\right|_{\{\phi_j\}\rightarrow \{\partial_{J_j}\}} - J_i
\end{equation}
while those associated with the action principle are

\begin{equation}
  \hat H_\alpha \equiv \partial_{g_\alpha} -
  \left.f_\alpha\right|_{\{\phi_j\}\rightarrow \{\partial_{J_j}\}} \; .
\end{equation}

Thus one obtains the commutation relations 

\begin{equation}
  [\hat L_i, \hat H_\alpha] = 0 \, , \qquad [\hat H_\alpha, \hat H_\beta] = 0 \; .
\end{equation}

Let us now write $\hat O_\alpha = \hat H_\alpha + \hat
\Delta_\alpha$. If $\hat L_i Z =0$ and $\hat O_\alpha Z=0$, then
$[\hat O_\alpha, \hat L_i]Z = 0$, or

\begin{equation} 
  \hat L_i \hat \Delta_\alpha Z = 0 \; .
\end{equation}

If $\hat\Delta_\alpha Z$ is non-zero, it is a solution of the Schwinger
Dyson equations. Moreover there is only one discrete possibility:
$\Delta_\alpha Z = c_\alpha\{g_\beta\} Z$, where $c_\alpha$ is some
function of the couplings,  so that $\hat H_\alpha Z = -
c_\alpha\{g_\beta\}Z$. The coefficients $c_\alpha$ can be absorbed
by a coupling constant dependent re-scaling $Z\rightarrow
Z^{\prime}= e^{\Omega\{g_\beta\}}Z$, where 

\begin{equation}
  \hat H_\alpha Z^\prime = [\hat H_\alpha, e^\Omega] Z - e^\Omega c_\alpha Z =
  \left((\partial_{g_\alpha}- c_\alpha )e^\Omega\right) Z = 0 \; .
\end{equation}

The existence of a solution to the equations $(\partial_{g_\alpha} -
c_\alpha)e^\Omega =0$ requires $\partial_{g_\alpha}c_\beta -
\partial_{g_\beta}c_\alpha =0$, which follows from $(\partial_{g_\alpha}c_\beta
- \partial_{g_\beta}c_\alpha)Z = -[\hat H_\alpha, \hat H_\beta] Z = 0$. The
re-scaled partition function satisfies both the Schwinger-Dyson and Schwinger
action principle equations, $\hat L_i Z^\prime = \hat H_\alpha Z^\prime =0$,
even though we started with just the Schwinger-Dyson equations.
\section{Non-Polynomial Actions}
Although we have focused on polynomial actions, our discussion of
the solution set of the Schwinger action principle equations can be
readily extended to non-polynomial actions. A simple non-polynomial
action is that one plaquette QED,  with action $S= \beta
\cos\theta$.  The generating functional

\begin{equation}
  \label{standgen}
  Z(J,\tilde J) = \int_{-\pi}^{\pi} d\theta e^{\beta\cos\theta
    + Je^{i\theta} + \tilde J e^{-i\theta}} \; ;
\end{equation}
is a solution of the differential equations

\begin{align}
  \label{SDP}
  &\left[\frac{\beta}{2}(\partial_J - \partial_{\tilde J}) + (J\partial_J
    -\tilde J \partial_{\tilde J})\right]Z(J,\tilde J) = 0 \\
  \label{PAP}
  &\left[\partial_\beta -\frac{1}{2}(\partial_J + \partial_{\tilde
      J})\right]Z(J,\tilde J) = 0\\
  \label{cons}
  &\partial_J \partial_{\tilde J} Z(J,\tilde J) = Z(J,\tilde J) \; ;
\end{align}
where \eqref{SDP} and \eqref{PAP} are the Schwinger-Dyson and action principle
equations, while \eqref{cons} is a constraint equation.  In fact these equations
have a two parameter (one if you neglect the normalization) class of solutions
given by linear combinations of basis solutions 

\begin{equation}
  Z(J,\tilde J) = \int_\Sigma d\theta e^{\beta\cos\theta + Je^{i\theta} + \tilde
    J e^{-i\theta}} \; ;
\end{equation}
for contours $\Sigma$ equivalent to either $\Sigma_1$ and $\Sigma_2$
in figure 9 (assuming real positive $\beta$).

\begin{center}
  \includegraphics{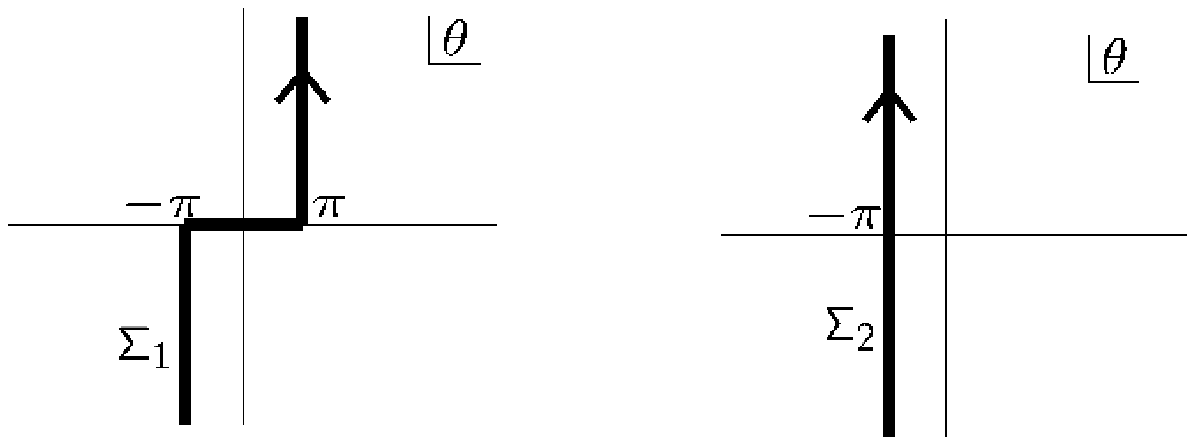}
  \bigskip

  \noindent{Figure 9.}
\end{center}

Note that integration over $\Sigma_1-\Sigma_2$ is equivalent to the
integral over the usual compact path $\theta =[-\pi,\pi]$.  The
possible physical relevance of the exotic solutions, upon
generalizing to a theory in a finite number of dimensions, is not
manifest as it was for polynomial theories.  In the former case,
symmetry breaking vacua were clearly related to exotic solutions.
Our experience with the polynomial theories leads us to  speculate
that the exotic solutions for lattice gauge theories are related to
physically realizable phases of gauge theory.  Certainly we expect
that the appearance of phase boundaries in gauge theories, via the
accumulation of Lee-Yang zeros, is closely related to the collapse
of the solution set in the thermodynamic limit.
\section{Conclusions and Outlook}
We have examined the complete set of solutions of the differential
equations which follow from the Schwinger action principle.  While
only one of these solutions corresponds to the usual path integral,
the other solutions, which involve complexified path integrals have
physical relevance.  On the one hand the manner in which the full
solution set collapses in the thermodynamic (or analogous
$g_T\rightarrow 0$) limit is related, via Stokes phenomena, to the
accumulation of Lee-Yang zeroes at phase boundaries. On the other
hand, exotic solutions may themselves be physical, with theta vacua
and symmetry breaking vacua as known examples.  In the zero
dimensional case, we have proven that Borel resummations of
perturbation series, having various singularity avoiding contours of
integration in the complex Borel variable, solve the action
principle equations and therefore correspond to various complexified
``path'' integrals.

While we have explicitly discussed the solution set of the Schwinger
Dyson and action principle equations for zero dimensional models,
one can readily generalize to lattice models in multi-dimensions, in
which case one finds a huge solution set.  The basis set of
solutions to the Schwinger-Dyson equations for a scalar field theory
on a lattice can be written in terms of the zero dimensional
solutions $Z^{(0)}(J)$ as follows;

\begin{equation}
  Z = \exp(K_{ij}\frac{\partial}{\partial J_i}\frac{\partial}{\partial
    J_j})\prod_k Z^{(0)}_k(J_k)
\end{equation}
where $K_{ij}$ is the lattice kinetic term, and the zero dimensional solution
$Z^{(0)}_k$ may be different at each lattice site $k$. For a polynomial
potential, the number of independent solutions grows exponentially with the
number of lattice sites. Like internal symmetries, space-time symmetries may be
broken by the choice of integration paths. Determining the collapse of the
solution set in the thermodynamic and continuum limits is a difficult problem.
It would be very interesting if exotic solutions with different integration
paths at different sites have physical relevance.
\section*{Acknowledgements}
We wish to thank Santiago Garcia for past collaboration related to the
present work. G.G. thanks D. Ferrante and C. Pehlevan for many enlightening
conversations.
\end{document}